\documentclass[12pt]{article}

\author{Yu.~M.~Zinoviev
       \thanks{E-mail address: yurii.zinoviev@ihep.ru} \\
               {\it Institute for High Energy Physics} \\
               {\it Protvino, Moscow Region, 142280, Russia}}
\title{On Dual Formulations \\
       of Massive Tensor Fields}

\date{}
 
\begin{document}

\maketitle

\begin{abstract}
In this paper we investigate dual formulations for massive tensor
fields. Usual procedure for construction of such dual formulations
based on the use of first order parent Lagrangians in many cases
turns out to be ambiguous. We propose to solve such ambiguity by using
gauge invariant description of massive fields which works both in
Minkowski space as well as (Anti) de Sitter spaces. We illustrate our
method by two concrete examples: spin-2 "tetrad" field $h_\mu{}^a$,
the dual field being "Lorentz connection" $\omega_\mu{}^{ab}$ and
"Riemann" tensor $R_{\mu\nu}{}^{ab}$ with the dual
$\Sigma_{\mu\nu}{}^{abc}$.
\end{abstract}

\thispagestyle{empty}
\newpage
\setcounter{page}{1}

\section*{Introduction}

Investigations of dual formulations for tensor fields are important
for understanding of alternative formulations of known theories like
gravity as well as understanding of their role in superstrings.
Common procedure for obtaining such dual formulations is based on the
parent first order Lagrangians. But such procedure turns out to be
simple and unambiguous for completely antisymmetric form fields only.
Dual formulations for more general massless fields where investigated
recently in \cite{BCH03,MV04,AHS04}. To illustrate the reason let us
compare two simplest cases: massive $s=1$ and $s=2$ fields. First
order formulation of massive spin-1 field requires two fields
$(F^{[\mu\nu]},A_\mu)$  treated as independent ones. The first order
Lagrangian has the form:
$$
{\cal L}_I = \frac{1}{4} F^{\mu\nu} F_{\mu\nu} - \frac{1}{2}
F^{\mu\nu} (\partial_\mu A_\nu - \partial_\nu A_\mu) + \frac{m^2}{2}
A_\mu{}^2
$$
If one solves the algebraic equation of motion for the $F^{\mu\nu}$
field and put result back into the Lagrangian one obtains usual
second order description for massive spin-1 particle in terms of
vector field $A_\mu$. But then the mass $m \ne 0$ the equation of
motion for vector field
$$
\frac{\delta {\cal L}}{\delta A_\mu} = m^2 A_\mu + (\partial F)_\mu =
0 $$
turns out to be algebraic too, so one can proceed by solving this
equation. Then putting the result back into the Lagrangian one gets
dual formulation of the same particle in terms of antisymmetric
tensor field $F^{\mu\nu}$:
$$
{\cal L}_{II} = - \frac{1}{2} (\partial F)_\mu{}^2 + \frac{m^2}{4}
F_{\mu\nu}{}^2
$$
where in order to get canonical normalization of kinetic term we make
a rescaling $F^{\mu\nu} \rightarrow m F^{\mu\nu}$. Note that the
kinetic term is invariant under the gauge transformations
$$
\delta F^{\mu\nu} = \varepsilon^{\mu\nu\alpha\beta} \partial_\alpha
\xi_\beta
$$
As a result in the massless limit $m \rightarrow 0$ such theory
describes massless spin-0 particle.

Now let us turn to the spin-2 case. The simplest and most common
description of such particle uses symmetric second rank tensor field
$h_{(\mu\nu)}$. But there is no any combination of first derivatives
for this field which will be invariant under the gauge transformations
$\delta h_{\mu\nu} = \partial_\mu \xi_\nu + \partial_\nu \xi_\mu$. So
to construct first order Lagrangian one has to abandon the symmetry
property and use "tetrad" field $h_\mu{}^a$ with modified gauge
transformations $\delta h_\mu{}^a = \partial_\mu \xi^a$. Then
introducing auxiliary field $\omega_\mu{}^{[ab]}$ one can construct
the following first order Lagrangian:
\begin{eqnarray*}
{\cal L}_I &=& - \frac{1}{2} \omega^{\mu,\alpha\beta} 
\omega_{\alpha,\mu\beta} + \frac{1}{2} \omega^\mu \omega_\mu -
\partial_\beta \omega^{\mu,\alpha\beta} h_{\alpha\mu} -
\partial^\alpha \omega^\mu h_{\mu\alpha} + (\partial \omega) h - \\
 && - \frac{m^2}{2} ( h^{(\mu\nu)} h_{(\mu\nu)} - h^2) +
 a h^{[\mu\nu]} h_{[\mu\nu]}
\end{eqnarray*}
Now we face the ambiguity in the mass terms. Indeed, if we solve the
algebraic equation of motion for the $\omega$ field and put result
back into the Lagrangian we obtain usual second order Lagrangian for
the symmetric part of $h_\mu{}^a$ while antisymmetric part turns out
to be non dynamical and completely decouples from it. But it is the
mass terms that determines the structure of kinetic terms in the dual
theory so that starting from different values for $a$ parameter we
will get different dual versions. For example, if one drops
antisymmetric part from the mass terms by choosing $a=0$ it will
persist in the kinetic terms serving as a Lagrangian multiplier and
giving differential constraint on $\omega$ field. If one will not
insist on the canonical form of the massless part in the first order
parent Lagrangian then even more arbitrary parameters could be
introduced \cite{CMU01,CMU02,CMU03}.

In this paper we are going to show that one of the way to resolve
such ambiguities is to use gauge invariant description of massive
tensor fields \cite{Zin01,Zin02,Zin03,Zin03a}. To illustrate how such
a procedure works, let us ones again use simplest spin-1 case. As is
well known gauge invariant description of massive spin-1 particle
requires introduction of additional Goldstone scalar field. Working
with first order formalism it is natural to use first order
description for all fields under consideration. So we introduce a pair
$(\pi^\mu, \varphi)$ and consider the following Lagrangian:
\begin{equation}
{\cal L} = \frac{1}{4} F_{\mu\nu}{}^2 - \frac{1}{2} F^{\mu\nu}
(\partial_\mu A_\nu - \partial_\nu A_\mu) - \frac{1}{2} \pi_\mu{}^2 +
\pi^\mu \partial_\mu \varphi - m \pi^\mu A_\mu
\end{equation}
It is easy to check that this Lagrangian is invariant under the gauge
transformations:
$$
\delta A_\mu = \partial_\mu \lambda \qquad \delta \varphi = m \lambda
$$
Moreover, if one solves algebraic equation of motion for $F^{\mu\nu}$
and $\pi^\mu$ fields and put the result back into the Lagrangian one
obtains usual second order formulation of massive spin-1 particle in
terms of vector $A_\mu$ and scalar $\varphi$ fields. Now let us try
to solve the equations for the $A_\mu$ and $\varphi$ fields instead.
First of all we face the fact that vector field $A_\mu$ enters the
Lagrangian only linearly working as a Lagrangian multiplier. Thus
it's equation
$$
\frac{\delta {\cal L}}{\delta A^\mu} = (\partial F)_\mu - m \pi_\mu =
0  $$
could be solved for the $\pi^\mu$ field and not for the field $A_\mu$
itself. Moreover the equation for the scalar field $\varphi$
$$ \frac{\delta {\cal L}}{\delta \varphi} = - (\partial \pi) = 0 $$
is not an independent one. Indeed it is easy to check that two
equations satisfy the relation:
$$
\partial^\mu \frac{\delta {\cal L}}{\delta A^\mu} - m 
\frac{\delta {\cal L}}{\delta \varphi} = 0
$$
which is just a consequence of the invariance under $\lambda$ gauge
transformations. So we can't express $\varphi$ field in terms of the
others, but it's not necessary because if put the solution of $A_\mu$
equation $\pi_\mu = - \frac{1}{m} (\partial F)_\mu$ into the initial
Lagrangian the scalar field $\varphi$ completely decouples leaving us
with:
$$
{\cal L} = - \frac{1}{2 m^2} (\partial F)_\mu{}^2 + \frac{1}{4}
F_{\mu\nu}{}^2
$$

In the following two sections we give two explicit examples of the
dual formulations for massive tensor fields obtained by the procedure
described above. The first one will be for the massive "tetrad" field
$h_\mu{}^a$ the dual field being "Lorentz connection"
$\omega_\mu{}^{ab}$. The second example will be the "Riemann" tensor
$R_{\mu\nu}{}^{ab}$ with the dual $\Sigma_{\mu\nu}{}^{abc}$.

\section{Massive spin 2}

Gauge invariant description of massive spin-2 particle requires
introduction of two Goldstone fields, namely the vector $A_\mu$ and
scalar $\varphi$ ones \cite{Zin01}. As we have already noted, it is
natural to use first order formulation for all fields under the
consideration, so we introduce three pairs of fields: $(h_\mu{}^a,
\omega_\mu{}^{ab})$, $(A_\mu, F^{ab})$ and $(\varphi, \pi^a)$
\cite{Zin03a}. Our starting point will be the sum of free massless
Lagrangians in flat Minkowski space:
\begin{eqnarray}
{\cal L}_0 &=& {\cal L}_0 (\omega_\mu{}^{ab}, h_\mu{}^a) +
{\cal L}_0 (F^{ab}, A_\mu) + {\cal L}_0 (\pi^a, \varphi) \\
{\cal L}_0 (\omega_\mu{}^{ab}, h_\mu{}^a) &=& \frac{1}{2}
\left\{ \phantom{|}^{\mu\nu}_{ab} \right\} 
 \omega_\mu{}^{ac} \omega_\nu{}^{bc} - \frac{1}{2}
 \left\{ \phantom{|}^{\mu\nu\alpha}_{abc} \right\}
 \omega_\mu{}^{ab} \partial_\nu h_\alpha{}^c \nonumber \\
{\cal L}_0 (F^{ab}, A_\mu) &=& \frac{1}{4} F_{ab}{}^2 -
\frac{1}{2} \left\{ \phantom{|}^{\mu\nu}_{ab}  \right\}
F^{ab} \partial_\mu A_\nu \nonumber \\
{\cal L}_0 (\pi^a, \varphi) &=& - \frac{1}{2} \pi_a{}^2 +
\left\{ \phantom{|}^\mu_a \right\} \pi^a \partial_\mu \varphi
\nonumber
\end{eqnarray}
Here
$$
\left\{ \phantom{|}^{\mu\nu}_{ab} \right\} = \delta_a{}^\mu
\delta_b{}^\nu - \delta_a{}^\nu \delta_b{}^\mu
$$
and so on. This Lagrangian is invariant under the following local
gauge transformations:
\begin{equation}
\delta_0 h_{\mu a} = \partial_\mu \xi_a + \eta_{\mu a} \qquad
\delta_0 \omega_\mu{}^{ab} = \partial_\mu \eta^{ab} \qquad
\delta_0 A_\mu = \partial_\mu \lambda
\end{equation} 

In (Anti) de Sitter spaces with nonzero cosmological constant gauge
invariance requires introduction of mass-like terms into the
Lagrangian as well as appropriate corrections for the gauge
transformation laws in much the same way as nonzero mass in Minkowski
space does. One of the pleasant features of gauge invariant
description of massive particles is the possibility to consider
general case of massive particle in (Anti) de Sitter as well as flat
spaces including all possible massless and partially massless
limits \cite{Zin01,Zin02}. Moreover, as was shown in \cite{MV04},
duality procedure for the massless fields in spaces with nonzero
cosmological constant works very similar to the one for massive fields
in flat space. Here we consider such a general case with nonzero mass
and cosmological constant. To simplify the formula we restrict
ourselves to space-times with the dimension $d=4$ only, but the whole
construction could be easily generalized to any dimensions $d \ge 3$.

Working with the first order formalism it is very convenient to 
use tetrad formulation of the underlying (Anti) de Sitter space.
We denote tetrad as $e_\mu{}^a$ (let us stress that it is not a
dynamical quantity here, just a background field) and Lorentz
covariant derivative as $D_\mu$. (Anti) de Sitter space is a 
constant curvature space with zero torsion, so we have:
\begin{equation}
D_{[\mu} e_{\nu]}{}^a = 0, \qquad [ D_\mu, D_\nu ] v^a =
 \kappa (e_\mu{}^a e_\nu{}^b - e_\mu{}^b e_\nu{}^a) v_b
\end{equation} 
where $\kappa = - 2 \Lambda/(d-1)(d-2) = - \Lambda/3$.

Now we replace all the derivatives in the Lagrangian and gauge
transformation laws by the covariant ones. Due to noncommutativity
of covariant derivatives the Lagrangian becomes non invariant
and we get:
\begin{equation}
\delta_0 {\cal L}_0 = 2\kappa \omega^a \xi_a - 2\kappa
h^{ab} \eta_{ab}
\end{equation} 
But the invariance could be restored by adding low derivative terms
to the Lagrangian:
\begin{eqnarray}
\Delta {\cal L} &=& \frac{m}{\sqrt{2}} \left[ 2 \omega^\mu A_\mu +
F^{\mu\nu} h_{\mu\nu} \right] + a_0 \pi^\mu A_\mu - \nonumber \\
 && - \frac{a_0{}^2}{6} ( h^{\mu\nu} h_{\nu\mu} - h^2) +
 \frac{m a_0}{\sqrt{3}} h \varphi + m^2 \varphi^2
\end{eqnarray}
as well as corresponding additional terms to the gauge transformation
laws:
\begin{eqnarray}
\delta' h_\mu{}^a &=& \frac{m}{\sqrt{2}} e_\mu{}^a \lambda \qquad
\delta' \omega_\mu{}^{ab} = - \frac{a_0{}^2}{6} (e_\mu{}^a \xi^b -
e_\mu{}^b \xi^a) \nonumber \\
\delta' F^{ab} &=& - m \sqrt{2} \eta^{ab} \qquad
\delta' A_\mu = \frac{m}{\sqrt{2}} \xi_\mu \\
\delta' \pi^a &=& \frac{m a_0}{\sqrt{2}} \xi^a \qquad
\delta' \varphi = - a_0 \lambda
\end{eqnarray}
Here $ a_0{}^2 = 3 m^2 + 6 \kappa $. The case $a_0 = 0$ corresponds
to the so called partial massless particles in the de Sitter space
and requires special treatment. In what follows we will assume that
$a_0 \ne 0$.

Now having in our disposal complete first order Lagrangian we try to
solve the equations for the $h_\mu{}^a$, $A_\mu$ and $\varphi$
fields. Let us start with the equation for $A_\mu$ field. Inspection
of the Lagrangian reveals that this field enters the Lagrangian only
linearly, so that it's equation
\begin{equation}
\frac{\delta {\cal L}}{\delta A_\mu} = D^\alpha F_{\alpha\mu} +
m \sqrt{2} \omega_\mu + a_0 \pi_\mu
\end{equation}
could not be solved for the $A_\mu$ field itself. But for $a_0 \ne 0$
it could be easily solved for $\pi^\mu$:
\begin{equation}
\pi_\mu = - \frac{1}{a_0} [ D^\alpha F_{\alpha\mu} + m \sqrt{2}
 \omega_\mu ]
\end{equation}
In this, if one put this result back into the Lagrangian, both
$A_\mu$ as well as $\pi_\mu$ drop out. Two other equations look like:
\begin{eqnarray}
\frac{\delta {\cal L}}{\delta h^{\mu\nu}} &=& R_{\nu\mu} - \frac{1}{6}
g_{\mu\nu} R + \frac{m}{\sqrt{2}} F_{\mu\nu} - \frac{a_0{}^2}{3}
(h_{\nu\mu} - g_{\mu\nu} h) + \frac{m a_0}{\sqrt{2}} g_{\mu\nu}
\varphi = 0 \nonumber \\
\frac{\delta {\cal L}}{\delta \varphi} &=& - (D \pi) + 
\frac{m a_0}{\sqrt{2}} h + 2 m^2 \varphi = 0
\end{eqnarray}
Here we introduce the tensor
$$
R_{\mu\nu}{}^{\alpha\beta} = D_\mu \omega_\nu{}^{\alpha\beta} -
D_\nu \omega_\mu{}^{\alpha\beta}
$$
as well as it's contractions $R_\mu{}^\alpha =
R_{\mu\nu}{}^{\alpha\nu}$ and $R = R_\mu{}^\mu$. Note that the
derivatives $D_\mu$ are covariant under the background Lorentz
connection only, so $R_{\mu\nu}{}^{\alpha\beta}$ is not truly gauge
invariant object. For example, under $\delta
\omega_\mu{}^{\alpha\beta} = D_\mu \eta^{\alpha\beta}$ we have
$\delta R_{\mu\nu} = - 2 \kappa \eta_{\mu\nu}$.
By taking a trace of the $h_{\mu\nu}$ equation
$$
g^{\mu\nu} \frac{\delta {\cal L}}{\delta h^{\mu\nu}} = - R + a_0{}^2
h + \frac{4 m a_0}{\sqrt{3}} \varphi = 0
$$
and comparing it with the $\varphi$ equation it is easy to see that
this equations are not independent but satisfy:
$$
D^\mu \frac{\delta {\cal L}}{\delta A^\mu} + \frac{m}{\sqrt{2}}
g^{\mu\nu} \frac{\delta {\cal L}}{\delta h^{\mu\nu}} - a_0
\frac{\delta {\cal L}}{\delta \varphi} = 0
$$
This last relation clearly shows that this dependency is just the
consequence of the  $\lambda$ gauge invariance. So we cannot solve
this equations for the $h_{\mu\nu}$ and $\varphi$ fields
simultaneously. But if one solves the $h_{\mu\nu}$ equation as:
\begin{equation}
h_{\mu\nu} = \frac{3}{a_0{}^2} [ R_{\nu\mu} - \frac{1}{6}
g_{\mu\nu} R - \frac{m}{\sqrt{2}} F_{\mu\nu} - \frac{m a_0}{3\sqrt{2}}
g_{\mu\nu} \varphi ]
\end{equation}
and put this expression into original Lagrangian (together with
expression for $\pi_\mu$) one sees that the scalar field $\varphi$
completely decouples. As a final result one obtains the second order
Lagrangian containing two fields only: the "gauge" field
$\omega_\mu{}^{\alpha\beta}$ and "Goldstone" field $F^{\alpha\beta}$:
\begin{eqnarray}
{\cal L}_{II} &=& \frac{1}{2} (R^{\mu\nu} R_{\nu\mu} - \frac{1}{3}
R^2) - \frac{1}{6} (D^\alpha F_{\alpha\mu})^2 + \nonumber \\
 && + \frac{m}{\sqrt{2}} (\omega^{\mu,\nu\alpha} D_\alpha F_{\mu\nu}
 + \frac{1}{3} \omega^\mu D^\alpha F_{\alpha\mu}) - \nonumber \\
  && - (\frac{m^2}{2} + \kappa) \omega^{\mu,\alpha\beta}
  \omega_{\alpha,\mu\beta} + (\frac{m^2}{6} + \kappa) \omega^\mu
  \omega_\mu + \frac{\kappa}{2} F_{\mu\nu}{}^2
\end{eqnarray}

This Lagrangian has general structure common for all gauge invariant
Lagrangians describing massive particles. It contains sum of the
kinetic terms for two fields $\omega_\mu{}^{\alpha\beta}$ and
$F^{\alpha\beta}$, cross terms with one derivative as well as mass
terms. (It could seems strange to have mass-like term for the
Goldstone field $F^{\mu\nu}$, but it's not unnatural in (Anti) de
Sitter spaces.) And indeed this Lagrangian is invariant under the
following gauge transformations:
\begin{equation}
\delta \omega_\mu{}^{\alpha\beta} = D_\mu \eta^{\alpha\beta} \qquad
\delta F^{\alpha\beta} = - m \sqrt{2} \eta^{\alpha\beta}
\end{equation}
Besides, as a remnant of the $\xi$-symmetry of the initial first
order Lagrangian, the second order Lagrangian is invariant also under
the local shifts:
\begin{equation}
\delta \omega_\mu{}^{\alpha\beta} = - e_\mu{}^\alpha \xi^\beta +
e_\mu{}^\beta \xi^\alpha
\end{equation}
This invariance could be easily checked if one uses that under such
transformations
$ \delta R_{\mu\nu} = 2 D_\mu \xi_\nu + g_{\mu\nu} (D \xi) $
and take into account useful identity:
$ D^\nu R_{\mu\nu} - \frac{1}{2} D_\mu R = - 2 \kappa \omega_\mu $.

\section{Massive $R_{[\mu\nu]}{}^{[ab]}$ tensor}

For gauge invariant description of appropriate massive 
particle one needs two additional Goldstone fields: $\Phi_{\mu\nu}{}^a$ 
and $h_\mu{}^a$ \cite{Zin02}. So to construct first order form of such
description we introduce three pairs of fields \cite{Zin03,Zin03a}:
$(\Sigma_{\mu\nu}{}^{abc}, R_{\mu\nu}{}^{ab})$, $(\Omega_\mu{}^{abc},
\Phi_{\mu\nu}{}^a)$ and $(\omega_\mu{}^{ab}, h_\mu{}^a)$. The sum of
flat space massless Lagrangians:
\begin{eqnarray}
{\cal L}_0 &=& {\cal L}_0 (\Sigma_{\mu\nu}{}^{abc}, R_{\mu\nu}{}^{ab})
+ {\cal L}_0 (\Omega_\mu{}^{abc}, \Phi_{\mu\nu}{}^a) +
{\cal L}_0 (\omega_\mu{}^{ab}, h_\mu{}^a) \nonumber \\
{\cal L}_0 (\Sigma_{\mu\nu}{}^{abc}, R_{\mu\nu}{}^{ab})
&=& - \frac{3}{8} \left\{
\phantom{|}^{\mu\nu\alpha\beta}_{abcd} \right\}
\Sigma_{\mu\nu}{}^{abe} \Sigma_{\alpha\beta}{}^{cde} + \frac{1}{4}
\left\{ \phantom{|}^{\mu\nu\alpha\beta\gamma}_{abcde} \right\}
\Sigma_{\mu\nu}{}^{abc} \partial_\alpha R_{\beta\gamma}{}^{de}  \\
{\cal L}_0 (\Omega_\mu{}^{abc}, \Phi_{\mu\nu}{}^a) &=& - \frac{3}{4}
\left\{\phantom{|}^{\mu\nu}_{ab} \right\}
\Omega_\mu{}^{acd} \Omega_\nu{}^{bcd} + \frac{1}{4}
\left\{\phantom{|}^{\mu\nu\alpha\beta}_{abcd} \right\}
\Omega_\mu{}^{abc} \partial_\nu \Phi_{\alpha\beta}{}^d \nonumber
\end{eqnarray}
where ${\cal L}_0 (\omega_\mu{}^{ab}, h_\mu{}^a)$ is the same as before
is invariant under the following gauge transformations:
\begin{eqnarray}
\delta_0 R_{\mu\nu}{}^{ab} &=& \partial_\mu \chi_\nu{}^{ab} -
\partial_\nu \chi_\mu{}^{ab} + \psi_{\mu,\nu}{}^{ab} -
\psi_{\nu,\mu}{}^{ab} \qquad \delta_0 \Sigma_{\mu\nu}{}^{abc} =
\partial_\mu \psi_\nu{}^{abc} - \partial_\nu \psi_\mu{}^{abc}
\nonumber \\
\delta_0 \Phi_{\mu\nu}{}^a &=& \partial_\mu z_\nu{}^a - \partial_\nu
z_\mu{}^a + \eta_{\mu\nu}{}^a \qquad
\delta_0 \Omega_\mu{}^{abc} = \partial_\mu \eta^{abc} \\
\delta_0 h_{\mu a} &=& \partial_\mu \xi_a + \eta_{\mu a} \qquad
\delta_0 \omega_\mu{}^{ab} = \partial_\mu \eta^{ab} \nonumber
\end{eqnarray} 
Then we proceed in the same way as in the previous example. Again
to simplify formula we restrict ourselves to spaces with dimension
$d = 5$ only (it's the minimal dimension where tensor
$R_{\mu\nu}{}^{ab}$ has physical degrees of freedom), but all the
results could be generalized to arbitrary $d \ge 5$ case.
First of all we go to (Anti) de Sitter space by changing all the
derivatives by covariant ones. Further we compensate resulting
noninvariance of the Lagrangian by adding additional low derivative
terms to the Lagrangian:
\begin{eqnarray}
{\cal L}_1 &=& \frac{m}{2} \left\{
\phantom{|}^{\mu\nu\alpha\beta}_{abcd}
\right\} \Sigma_{\mu\nu}{}^{abc} \Phi_{\alpha\beta}{}^d - \frac{3m}{2}
\left\{ \phantom{|}^{\mu\nu\alpha}_{abc} \right\}
R_{\mu\nu}{}^{ad} \Omega_\alpha{}^{bcd} + \nonumber \\
 && + \frac{a_0}{2}
\left\{ \phantom{|}^{\mu\nu}_{ab} \right\} \Omega_\mu{}^{abc} 
h_\nu{}^c+ \frac{a_0}{2} \left\{ \phantom{|}^{\mu\nu\alpha}_{abc}
\right\} \omega_\mu{}^{ab} \Phi_{\nu\alpha}{}^c  - \nonumber \\
&& - \frac{a_0{}^2}{8} \left\{ \phantom{|}^{\mu\nu\alpha\beta}_{abcd} 
\right\}R_{\mu\nu}{}^{ab} R_{\alpha\beta}{}^{cd} - m a_0
\left\{ \phantom{|}^{\mu\nu\alpha}_{abc} \right\} R_{\mu\nu}^{ab}
h_\alpha{}^c - 3 m^2 \left\{ \phantom{|}^{\mu\nu}_{ab} \right\}
h_\mu{}^a h_\nu{}^b
\end{eqnarray} 
as well as corresponding terms to the gauge transformations:
\begin{eqnarray}
\delta_1 R_{\mu\nu}{}^{ab} &=& - \frac{m}{4} e_{[\mu}{}^{[a} 
z_{\nu]}{}^{b]} \qquad \delta_1 h_\mu{}^a = 2 a_0 z_\mu{}^a \nonumber \\
\delta_1 \Sigma_{\mu\nu}{}^{abc} &=& \frac{m}{8}
e_{[\mu}{}^{[a} \eta_{\nu]}{}^{bc]} + \frac{a_0{}^2}{3} e_{[\mu}{}^{[a}
\chi_{\nu]}{}^{bc]} - \frac{m a_0}{3} e_{[\mu}{}^{[a} e_{\nu]}{}^b
\xi^{c]}  \nonumber \\
\delta_1 \Phi_{\mu\nu}{}^a &=& 2 m (\chi_{\mu,\nu}{}^a -
\chi_{\nu,\mu}{}^a) + \frac{a_0}{6} e_{[\mu}{}^a \xi_{\nu]}  \\
\delta_1 \Omega_\mu{}^{abc} &=& - 4 m \psi_\mu{}^{abc} + \frac{a_0}{3}
e_\mu{}^{[a} \eta^{bc]} \nonumber \\
\delta_1 \omega_\mu{}^{ab} &=& - a_0 \eta_\mu{}^{ab} +
\frac{a_0{}^2}{3} \chi_\mu{}^{ab} - \frac{m a_0}{3} e_\mu{}^{[a}
 \xi^{b]} \nonumber
\end{eqnarray} 
Here $ a_0{}^2  = 6 m^2 + 3 \kappa $. As in the previous case $a_0 =
0$ corresponds to partially massless particle which require special
treatment. In what follows we will assume that $a_0 \ne 0$.

It will be convenient to introduce Lorentz covariant "field strength"
$$
\Sigma_{\mu\nu\alpha}{}^{abc} = D_{[\mu} \Sigma_{\nu\alpha]}{}^{abc}
$$
Because it's covariant under the background Lorentz connection only
it's not fully gauge invariant\footnote{Here and further on we use
simple "first Greek --- first Latin" rool for the contractions of
indices.}:
\begin{equation}
\delta \Sigma_{\mu\nu}{}^{abc} = D_{[\mu} \psi_{\nu]}{}^{abc} \quad
\Longrightarrow \quad \delta \Sigma_{\mu\nu}{}^{ab} = - \kappa
(\psi_{\mu,\nu}{}^{ab} - \psi_{\nu,\mu}{}^{ab})
\end{equation}
Let us give here useful identities:
\begin{eqnarray}
(D \Sigma)_{\mu\nu}{}^a - D_\mu \Sigma_\nu{}^a + D_\nu \Sigma_\mu{}^a 
&=& \kappa (\Sigma_{\mu,\nu}{}^a - \Sigma_{\nu,\mu}{}^a) \nonumber \\
(D \Sigma)_\mu - \frac{1}{2} D_\mu \Sigma &=& \frac{\kappa}{2}
\Sigma_\mu
\end{eqnarray}
Also we introduce a convenient linear combination:
$$
\hat{\Sigma}_{\mu\nu}{}^{ab} = \Sigma_{\mu\nu}{}^{ab} - \frac{1}{4}
e_{[\mu}{}^{[a} \Sigma_{\nu]}{}^{b]} + \frac{1}{18} \Sigma 
e_\mu{}^{[a} e_\nu{}^{b]}
$$
One of the reason why we choose this particular coefficients is a
simple form of transformations for this object, for example:
\begin{equation}
\delta \Sigma_{\mu\nu}{}^{abc} = e_{[\mu}{}^{[a} \chi_{\nu]}{}^{bc]}
\quad \Longrightarrow \quad \delta \hat{\Sigma}_{\mu\nu}{}^{ab} = - 
D_{[\mu} \chi_{\nu]}{}^{ab}
\end{equation}

Analogously, we introduce field strength for the $\Omega$ field:
$$
\Omega_{\mu\nu}{}^{abc} = D_\mu \Omega_\nu{}^{abc} - D_\nu
\Omega_\mu{}^{abc}
$$
It is also not a truly gauge invariant object
\begin{equation}
\delta \Omega_\mu{}^{abc} = D_\mu \eta^{abc} \quad \Longrightarrow
\quad \delta \Omega_\mu{}^{ab} = 2 \kappa \eta_\mu{}^{ab}
\end{equation}
The following identities:
\begin{eqnarray}
(D \Omega)_{\mu\nu}{}^{ab} - D_\mu \Omega_\nu{}^{ab} + D_\nu
\Omega_\mu{}^{ab} &=& - \kappa (\Omega_{\mu,\nu}{}^{ab} -
\Omega_{\nu,\mu}{}^{ab}) + \kappa e_{[\mu}{}^{[a} \Omega_{\nu]}{}^{b]}
\nonumber \\
(D \Omega)_\mu{}^a - \frac{1}{2} D_\mu \Omega^a &=& - 2 \kappa
\Omega_\mu{}^a
\end{eqnarray}
will be useful as well as convenient combination:
\begin{equation}
\hat{\Omega}_\mu{}^{ab} = \Omega_\mu{}^{ab} - \frac{1}{6} e_\mu{}^{[a}
\Omega^{b]}
\end{equation}

Now we are ready to construct dual formulation for this model. Our
task here to solve equations for $R_{\mu\nu}{}^{ab}$,
$\Phi_{\mu\nu}{}^a$ and $h_\mu{}^a$ fields. Let us start from the
$\Phi_{\mu\nu}{}^a$ equation. Once again we face that this field
enters the Lagrangian only linearly, so it's equation couldn't be
solved for the field $\Phi_{\mu\nu}{}^a$ itself. But for the case
$a_0 \ne 0$ it can be solved for the $\omega_\mu{}^{ab}$ field giving:
\begin{equation}
\omega_\mu{}^{ab} = \frac{1}{a_0} \left[ - \frac{3}{2} 
\hat{\Omega}_\mu{}^{ab} -  6 m \Sigma_\mu{}^{ab} + m e_\mu{}^{[a}
\Sigma^{b]} \right]
\end{equation}
Hence both $\Phi_{\mu\nu}{}^a$ as well as $\omega_\mu{}^{ab}$ field
drop out from the resulting second order Lagrangian. Now let us turn
to the equation for $R_{\mu\nu}{}^{ab}$ and $h_\mu{}^a$ fields.
Comparing the trace of $R_{\mu\nu}{}^{ab}$ equation with the
$h_\mu{}^a$ one it is not hard to check that they are not independent
and satisfy:
$$
2 m \delta_d{}^b \frac{\delta {\cal L}}{\delta R_{ab}{}^{cd}} - a_0
\frac{\delta {\cal L}}{\delta h_a{}^c} = D_b 
\frac{\delta {\cal L}}{\delta \Phi_{ab}{}^c}
$$
which is a simple consequence of $z_\mu{}^a$ gauge invariance. So it
is not possible to express both $R_{\mu\nu}{}^{ab}$ and $h_\mu{}^a$
in terms of the other fields simultaneously. We proceed by solving
equation for $R_{\mu\nu}{}^{ab}$:
\begin{equation}
R_{\mu\nu}{}^{ab} = - \frac{1}{a_0{}^2} \left[ 3 
\hat{\Sigma}_{\mu\nu}{}^{ab} + \frac{3 m}{2} (\Omega_{\mu,\nu}{}^{ab}
- \Omega_{\nu,\mu}{}^{ab}) + \frac{m a_0}{2} e_{[\mu}{}^{[a}
h_{\nu]}{}^{b]} \right]
\end{equation}
Now we put this expression into the initial first order Lagrangian.
In this, $h_\mu{}^a$ field completely decouples (and that serves as a
check for rather lengthy calculations). As a final result we obtain a
second order Lagrangian containing two fields: "gauge" field
$\Sigma_{\mu\nu}{}^{abc}$ and "Goldstone" field $\Omega_\mu{}^{abc}$:
\begin{eqnarray}
\frac{a_0{}^2}{9} {\cal L}_{II} &=& \frac{1}{2}
(\hat{\Sigma}_{ab}{}^{cd}
\hat{\Sigma}_{cd}{}^{ab} - 4 \hat{\Sigma}_a{}^b \hat{\Sigma}_b{}^a +
\hat{\Sigma} \hat{\Sigma}) - \frac{1}{2} (\omega^{a,bc} \omega_{b,ac} -
\omega^a \omega_a) + \nonumber \\
&& + m(\hat{\Sigma}_{ab}{}^{cd} \Omega_{c,d}{}^{ab} +
2 \hat{\Sigma}_a{}^b \Omega_b{}^a) - \frac{\kappa}{4} (\Omega^{a,bcd}
 \Omega_{c,dab} - \Omega^{ab} \Omega_{ba})  \nonumber \\
 && - \frac{a_0{}^2}{6} ( \Sigma^{ab, cde} \Sigma_{cd,eab} - 4 
\Sigma^{a,bc} \Sigma_{b,ac} + \Sigma^a \Sigma_a) 
\end{eqnarray}
This Lagrangian also has general structure common to all gauge
invariant Lagrangians describing massive particles: it contains a sum
of the kinetic terms for two fields, cross terms with one derivative
and mass terms. And indeed this Lagrangian is invariant under the
following gauge transformations:
\begin{equation}
\delta \Sigma_{\mu\nu}{}^{abc} = D_\mu \psi_\nu{}^{abc} - D_\nu 
\psi_\mu{}^{abc} \qquad \delta \Omega_\mu{}^{abc} = 4 m
\psi_\mu{}^{abc} 
\end{equation}
Bur this time a Goldstone field $\Omega_\mu{}^{abc}$ is a one form
being simultaneously a gauge field having it's own gauge invariance:
\begin{equation}
\delta \Omega_\mu{}^{abc} = D_\mu \eta^{abc} \qquad
\delta \Sigma_{\mu\nu}{}^{abc} = 0  
\end{equation}
At last, as a remnant of $\chi$ invariance of the initial first order
Lagrangian, second order Lagrangian is invariant under the local
shifts:
\begin{equation}
\delta \Sigma_{\mu\nu}{}^{abc} = e_{[\mu}{}^{[a} \chi_{\nu]}{}^{bc]}
\qquad \delta \Omega_\mu{}^{abc} = 0 
\end{equation}

\section*{Conclusion}

In this paper we have shown that gauge invariant description of
massive particles allows one to resolve ambiguities arising in the
construction of dual formulations using first order parent
Lagrangians. In this, one can easily consider general case with
nonzero masses as well as cosmological term. Note that the resulting
second order dual Lagrangians also turn out to be gauge invariant.
We restrict ourselves by considering two concrete examples, but it's
evident that such a procedure could be easily generalized for other
cases as well. At the same time, there are interesting models like
those describing partially massless particles that requires special
treatment and should be considered separately.

\end{document}